\documentclass[twocolumn,prl]{revtex4}
\usepackage{graphics,graphicx}
\begin{document}
\title{Quantum magnetic properties and metal-to-insulator transition in chemically doped calcium ruthenate perovskite}
\author{Deepak K. Singh$^{1,*}$}
\author{Arthur Ernst$^{2,3}$}
\author{Vitalii Dugaev$^{4}$}
\author{Yiyao Chen$^{1,5}$}
\author{Jagath Gunasekera$^{1,6}$}
\affiliation{$^{1}$Department of Physics and Astronomy, University of Missouri, Columbia, MO, USA}
\affiliation{$^{2}$Max-Planck-Institut f\"ur Mikrostrukturphysik, Weinberg 2, 06120 Halle, Germany}
\affiliation{$^{3}$Institut f\"ur Theoretische Physik, Johannes Kepler Universit\"at, 4040 Linz, Austria}
\affiliation{$^{4}$Department of Physics and Medical Engineering, Rzesz\'ow University of Technology, Rzesz\'ow, Poland}
\affiliation{$^{5}$Suzhou Institute of Nano-Tech and Nano-Bionics, Chinese Academy of Sciences, China}
\affiliation{$^{6}$Department of Mechanical and Mechatronic Engineering, Southern Illinois University, Edwardsville, IL, USA}
\affiliation{$^{*}$email: singhdk@missouri.edu}

\begin{abstract}
Ruthenates provide comprehensive platform to study a plethora of novel properties, such as quantum magnetism, superconductivity and magnetic fluctuation mediated metal-insulator transition. In this article, we provide an overview of quantum mechanical phenomenology in calcium ruthenium oxide with varying compositions. While the stochiometric composition of CaRuO$_{3}$ exhibits non-Fermi liquid behavior with quasi-criticality, chemically doped compounds depict prominent signatures of quantum magnetic fluctuations at low temperature that in some cases are argued to mediate in metal-insulator transition. In the case of cobalt doped- CaRuO$_{3}$, an unusual continuum fluctuation is found to persist deep inside the glassy phase of the material. These observations reflect the richness of ruthenate research platform in the study of quantum magnetic phenomena of fundamental importance.
\end{abstract}

\maketitle

Perovskite magnetic materials are at the frontline of solid state physics research. Experimental and theoretical exploration of perovskite magnets have resulted in numerous novel discoveries of both fundamental and practical importance.\cite{Varma} Some of the notable examples include the depiction of high temperature superconductivity,\cite{Bednorz} colossal magnetoresistance,\cite{Ramirez} high efficiency electrodes \cite{Yu} and quantum magnetic phenomena.\cite{Bishop,Balents,Coldea} Among various perovskite magnets, ruthenates are of special importance as a host of physical and magnetic properties are discovered in materials with varying stoichiometry.\cite{Deng} Calcium ruthenium oxide with 1-1-3 phase composition is one of the most studied yet not well understood compound.\cite{Kikugawa,Longo,Cao2,Cava,Mazin,Cao1,Khalifah,Tripathi,Felner,Mukuda} It is a non-Fermi liquid metal on the verge of magnetic instability.\cite{Kostic,Lee,Cao1,Klein} Chemical substitution on Ca or Ru site is known to cause metal-insulator transition with complex magnetic properties, including spin glass and spin liquid phenomena.\cite{Gunasekera1,Cao2,Gunasekera2,Gunasekera3,Gunasekera4,Chen} Here, we briefly discuss electrical, magnetic and spin-spin correlation properties in stoichiometric CaRuO$_{3}$ and some of the chemically doped versions.

\begin{figure}
\centering
\includegraphics[width=9. cm]{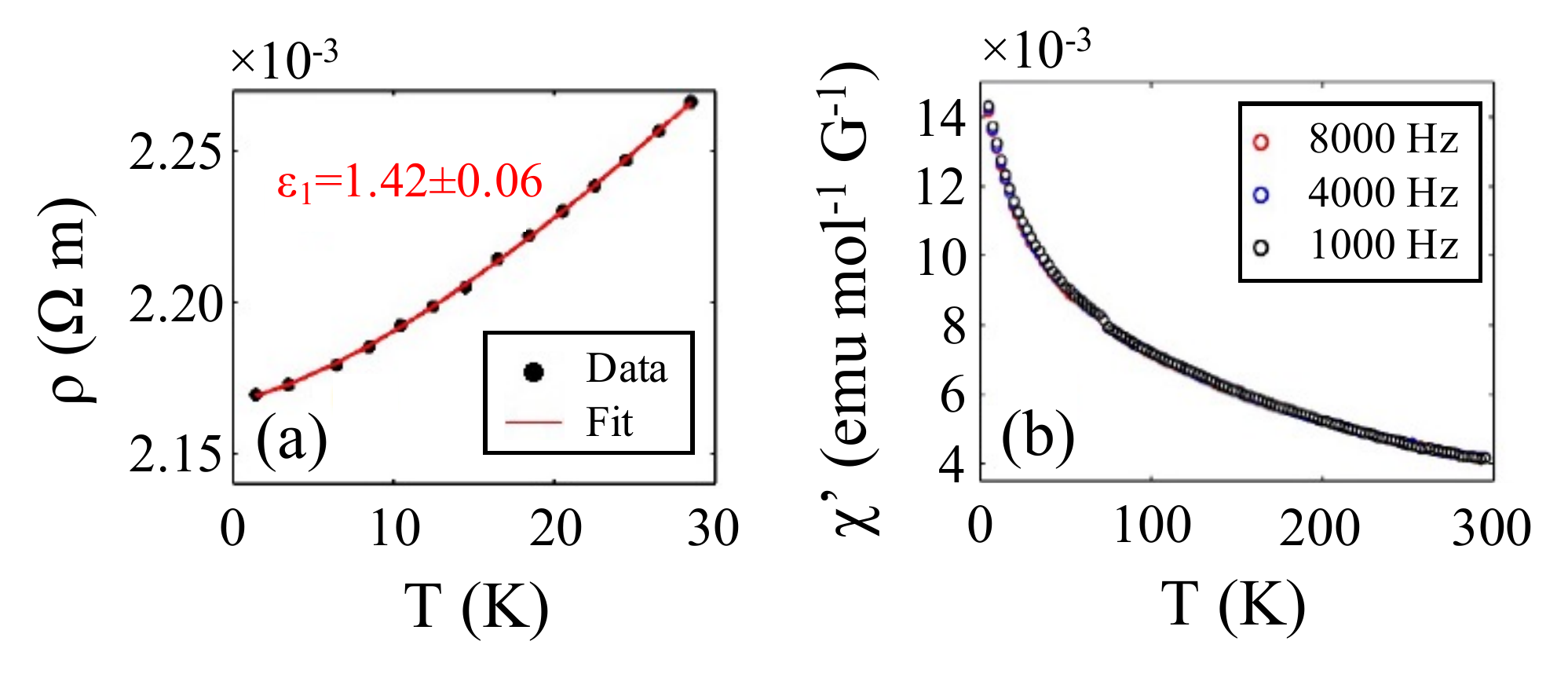} \vspace{-6mm}
\caption{(a) Electrical resistivity as a function of temperature in CaRuO$_{3}$, depicting non-Fermi-liquid state at low temperature. The resistivity data is fitted using the power-law equation, $\rho$ = $T$$^{\epsilon}$, with fitted value of $\epsilon$ = 1.42(0.06). (b) Static susceptibilities at few characteristic frequencies are plotted as a function of temperature. The static susceptibility exhibits paramagnetic behavior, which is ac frequency independent. 
} \vspace{-6mm}
\end{figure}

\textbf{Electrical and magnetic properties of CaRuO$_{3}$}: The perovskite compound crystallizes in the orthorhombic structure (Pnma crystallographic group) with lattice parameters of $a$ = 5.545 $\AA$, $b$=7.673 $\AA$ and $c$ =5.398 $\AA$.\cite{Cava,Cao1,Kikugawa,Gunasekera2,Akaogi} The octahedral crystalline electric field splits the fivefold degeneracy of Ru 4$d$$^{4}$ into ($t_{2g}$$^{4}$ and empty $e_g$, thus resulting in a low-spin electronic configuration.\cite{Mazin,Cao2} There are two peculiar characteristics attested to CaRuO$_{3}$: magnetic instability (close to magnetic stability) and non-Fermi liquid behavior. Unlike the ferromagnetic Sr counterpart SrRuO$_{3}$,\cite{Kikugawa} CaRuO$_{3}$ is non-magnetic.\cite{Gunasekera2} Detailed investigation of underlying magnetism using NMR measurements revealed a highly dynamic ground state with fluctuating Ru spins at low temperature.\cite{Yoshimura} Similarly, previous electrical measurements on polycrystalline and crystalline specimen suggested a non-Fermi liquid state in the system.\cite{Cao1} We performed detailed electrical and magnetic measurements on high quality polycrystalline CaRuO$_{3}$ to further understand the unusual behavior.\cite{Gunasekera2} Details about the sample synthesis can be found elsewhere.\cite{Gunasekera2} We show electrical and magnetic measurements data in Fig. 1. As we can see in Fig. 1a, the experimental data are well described by a power-law expression, $\rho \propto$ $T$$^{\epsilon}$ . The fitting parameter $\epsilon$ is found to exhibit non-integer value of 1.42(0.06) below $T$ $\sim$ 30 K; illustrating a non-Fermi-liquid (NFL) state at low temperature. This is consistent with previous observations on the high quality specimens of CaRuO$_{3}$. Electrical resistivity is also found to be independent of magnetic field application, suggesting a robust NFL state. The plot of real part of magnetic susceptibility $\chi$$^{’}$ in Fig. 1b reveals the paramagnetic behavior with a small cusp around $T$ $\sim$ 80 K. A similar phenomenon was previously attributed to the onset of a long-range magnetic order.\cite{Felner} However, no signature of any magnetic order was observed in elastic neutron scattering measurements.\cite{Gunasekera2} Bulk static susceptibilities are also found to be independent of ac measurement frequency, hence ruling out the occurrence of the spin glass state in the system. 

\begin{figure}
\centering
\includegraphics[width=8.8 cm]{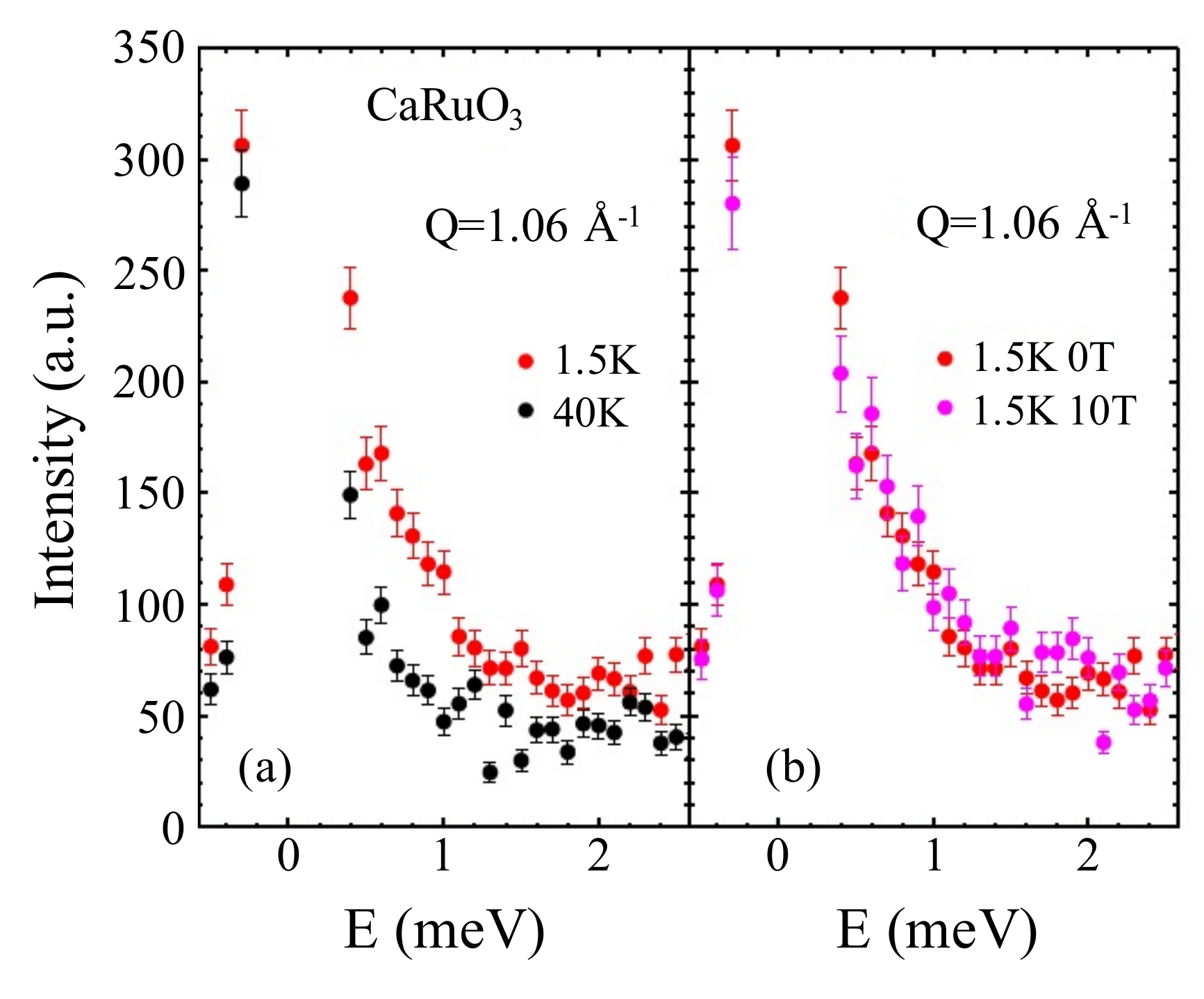} \vspace{-6mm}
\caption{Plot of inelastic neutron scattering intensity as a function of energy transfer $E$ ($E_i$ - $E_f$) at (a) two different temperatures of $T$ = 1.5 K and 40 K in zero magnetic field, and (b) at two different magnetic fields of $H$ = 0 T and 10 T at $T$ = 1.5 K. Inelastic measurements were performed at fixed $Q$ of 1.06 $\AA$$^{-1}$. Additional spectral weight is detected below $E$ =1 meV in zero field measurement. Inelastic data remains unaffected to magnetic field application.
} \vspace{-6mm}
\end{figure}

The ground state of CaRuO$_{3}$ is still a puzzle. No signature of antiferromagnetic or ferromagnetic order has ever been detected in this system in neutron scattering or magnetic measurements. Experimental investigation of dynamic behavior using inelastic neutron scattering measurements has revealed low energy excitation of short-range correlated spins at low temperature, see Fig. 2. The dynamic correlation of Ru$^{4+}$ ions do not exhibit any magnetic field dependence as no change in spectral weight is detected upto $H$ = 10 T of magnetic field application (Fig. 2b). It suggests a highly robust nature of magnetic fluctuation in CaRuO$_{3}$, also consistent with previous NMR measurements.\cite{Yoshimura} Previously, we investigated the spin fluctuation rate and its temperature dependence using the quantitative analysis of inelastic neutron spectra.\cite{Gunasekera2} In a first observation, a quasi-critical state, characterized by the divergence of relaxation time and short-range critical scaling of dynamic susceptibility, was inferred to persist at low temperature. Perhaps. the prevalent quantum mechanical spin dynamics prohibits the development of a static order in this compound. Clearly, the ground state in CaRuO$_{3}$ is more exciting than previously understood.

\begin{figure*}
\centering
\includegraphics[width=18 cm]{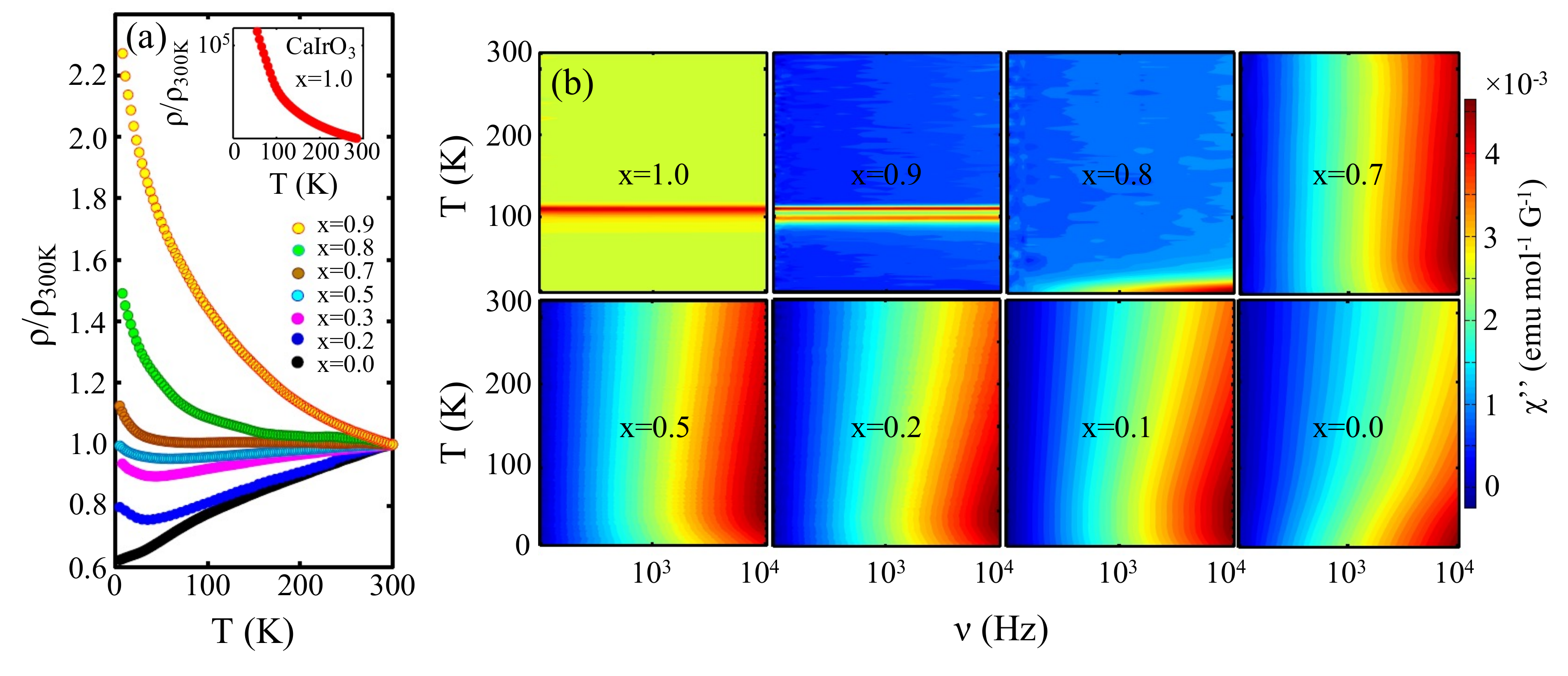} \vspace{-6mm}
\caption{(a) Metal-to-insulator transition as a function of x in Ca(Ir$_{x}$Ru$_{1-x}$)O$_{3}$. As x increases above 0.7, insulating behavior starts dominating. At x = 1, the compound CaIrO3 is a strong Mott insulator. (b) Temperature vs. frequency plots of dynamic susceptibility, $\chi$$^{”}$ for several chemical substitution coefficient x. For x $\leq$ 0.7, strong frequency-dependence of the dynamic susceptibilities are clearly observable, indicating significant magnetic fluctuations in the system. Onset of magnetic fluctuation coincides with the onset of metallic behavior in Ca(Ir$_{x}$Ru$_{1-x}$)O$_{3}$.} \vspace{-4mm}
\end{figure*}

\textbf{Magnetic fluctuation mediated metal-to-insulator transition in chemically substituted Ca($M$$_{x}$Ru$_{1-x}$)O$_{3}$, $M$ = Ir, Co}

Metal-insulator transition (MIT) in magnetic materials has been a subject of intense study.\cite{Imada} The transition is often accompanied by a change in the underlying magnetic characteristics, namely the transition from ferromagnetic to antiferromagnetic states or the onset of magnetic fluctuations in metallic state.\cite{Imada,Gunasekera1} In strongly correlated electron systems, the metal-insulator transition is argued to transition from the Hubbard model to Heisenberg regime where electrical conductivity is found to exhibit activated insulating behavior. Understanding the role of magnetic fluctuation in MIT is also enticing from the perspective of high temperature superconductors where the mechanism is argued to facilitate the Cooper pair formation.\cite{PLee} We have investigated this problem in 113-ruthenate perovskite. For this purpose, Ru ions are substituted by Ir and Co ions in CaRuO$_{3}$ compound. 

\textbf{A. Study of Ca(Ir$_{x}$Ru$_{1-x}$)O$_{3}$}: Since both CaRuO$_{3}$ and CaIrO$_{3}$ crystallize in the orthorhombic structure and manifest empty $e_g$-band electronic configurations of 4d$^{4}$ (near half-filled) and 5d$^{5}$ (half-filled) states in Ru$^{4+}$ and Ir$^{4+}$ ions, respectively, the experimental efforts provide direct insight into this problem.\cite{Gunasekera3} Moreover, CaIrO$_{3}$ is known to manifest antiferromagnetic order below $T_N$ $\sim$ 110 K.\cite{Sala2,Gunasekera1} Thus, the end compounds represent two diverse magnetic systems. Substitution of Ru ion by Ir ion induces a metal-to-insulator transition. The electrical transport measurement data are shown in Fig. 3a. Here, the normalized electrical resistivity ($\rho$/$\rho$$_{300K}$), where $\rho$$_{300K}$ is the resistivity at $T$ = 300 K) is plotted as a function of temperature for various chemical substitution coefficient x. As we can see, a gradual evolution from the metallic state, x = 0, to strongly insulating state, x = 1, is detected as the coefficient x varies from 0 to 1. The paramagnetic characteristic of CaRuO$_{3}$ prevails across the low chemical substitution of Ru by Ir. 

To understand the metal-to-insulator transition process, we have also performed detailed dynamic susceptibility measurements as a function of temperature and ac frequency for various substitution coefficients. The ac frequency varies from 10$^{2}$ Hz to 10$^{4}$ Hz, which corresponds to magnetic fluctuation time scale of $t \geq$ 100 $\mu$s. Experimental results are presented in the form of contour plots of dynamic susceptibilities in Fig. 3b.  In this figure, we see that at x = 1, the plot is characterized by a singular line at $T \simeq$ 110 K, indicating the absence of dynamic behavior. At x = 1, the compound CaIrO$_{3}$ is a strong antiferromagnetic Mott insulator with no obvious dynamic properties at long time scale. As x decreases below 0.7, the small regime of large $\chi$$^{"}$ at high frequency and low temperature extends to higher temperature, thus illustrating significant spin dynamics in compounds with coefficient x $<$ 0.7. In fact, the dynamic response to the ac susceptibility measurements is found to be strongest in x = 0.3 composition. Direct comparison of contour plots with electrical transport data (Fig. 3a) reveals one-to-one correspondence between the development of magnetic fluctuation and the onset of metallic behavior. At x = 0, the compound CaRuO$_{3}$ is a non-Fermi liquid metal with strongly dynamic magnetic properties, characterized by fluctuating Ru spins. The synergistic experimental investigation of electrical and magnetic properties suggests the presence of magnetic fluctuations in various compositions of Ca(Ir$_{x}$Ru$_{1-x}$)O$_{3}$ perovskites.\cite{Gunasekera3}

\textbf{B. Metal-insulator transition in Ca(Co$_{x}$Ru$_{1-x}$)O$_{3}$}: While the study of Ca(Ir$_{x}$Ru$_{1-x}$)O$_{3}$ has provided valuable insight in understanding the novel metal-insulator transition phenomena in 113-ruthenate perovskite, theoretical calculations hinted that substitution of Ru ion by isoelectronic Co-ion can lead to the formation of artificial spin-1/2 state in Ca(Co$_{x}$Ru$_{1-x}$)O$_{3}$.\cite{Gunasekera4} Moreover, the non-stoichiometric composition of calcium ruthenate also manifest strong metal-insulator transition process, see Fig. 4. Detailed study of electrical and magnetic properties of Ca(Co$_{x}$Ru$_{1-x}$)O$_{3}$ showed that the metal-insulator transition is comprised of both Hubbard and Heisenberg formalisms due to band filling.\cite{Imada} At low substitution coefficient of x $<$ 0.15, the system still exhibits weak conducting characteristics- perhaps on the verge of becoming an insulator. Above x = 0.15, the electrical resistance is described by the Arrhenius formula, thus exhibits activated behavior that are typically found in insulating materials. The localized Co-moments are described by Heisenberg formulation. Interestingly, the Co-Co interaction dominates the Ru-Co or Ru-Ru interaction in Ca(Co$_{x}$Ru$_{1-x}$)O$_{3}$.\cite{Gunasekera4} 

\begin{figure}
\centering
\includegraphics[width=8.8 cm]{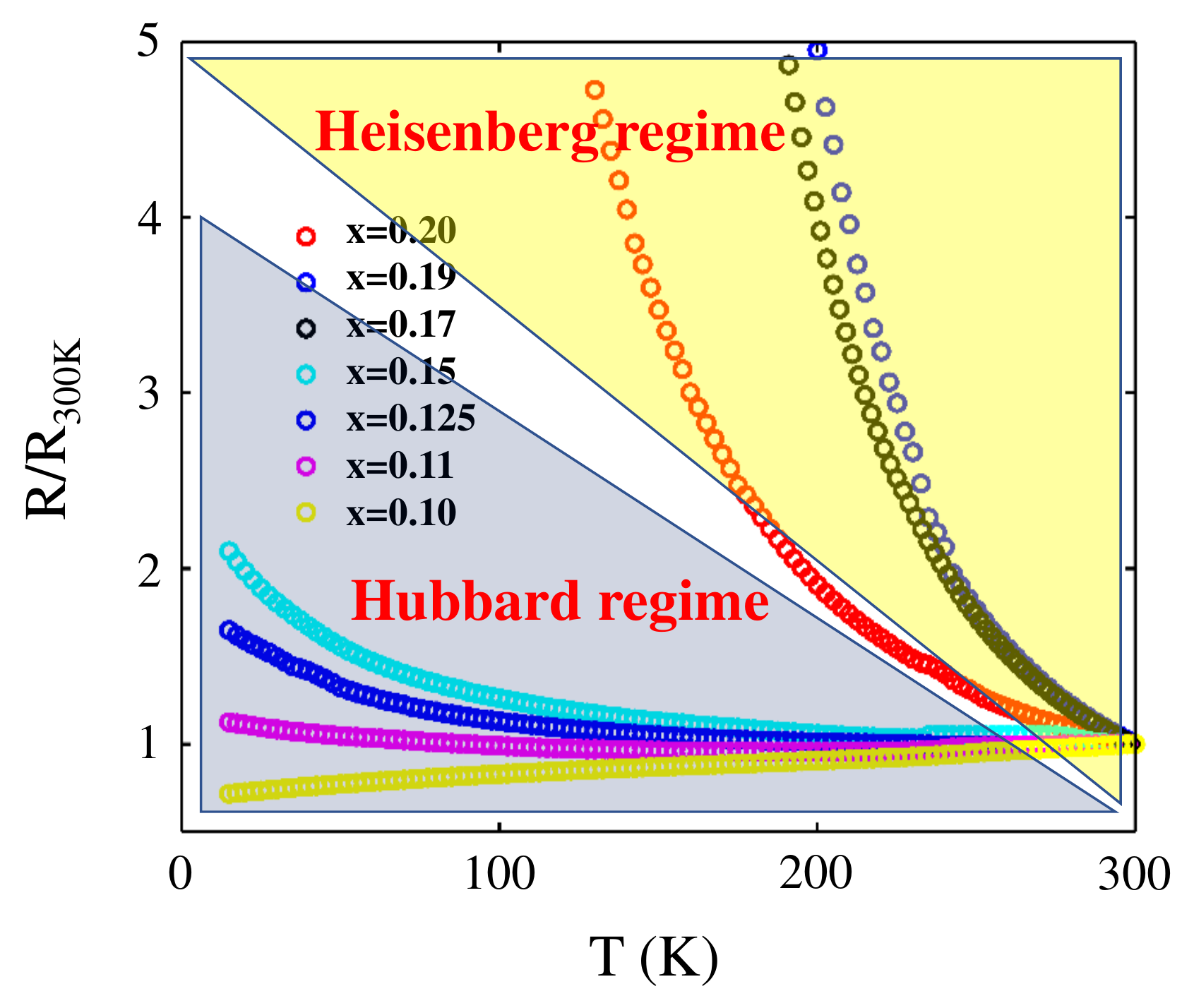} \vspace{-6mm}
\caption{Metal-insulator transition in Co-doped CaRuO$_{3}$. Here we show the plot of electrical resistivity versus temperature for various substitution coefficient x. As x increases, a transition from metal to insulator phase is clearly observed. At x$<$ 0.15, the system is on the verge of becoming an insulator. Above x = 0.15, plot of electrical resistivity suggests insulating characteristics.} \vspace{-6mm}
\end{figure}

\begin{figure*}
\centering
\includegraphics[width=17 cm]{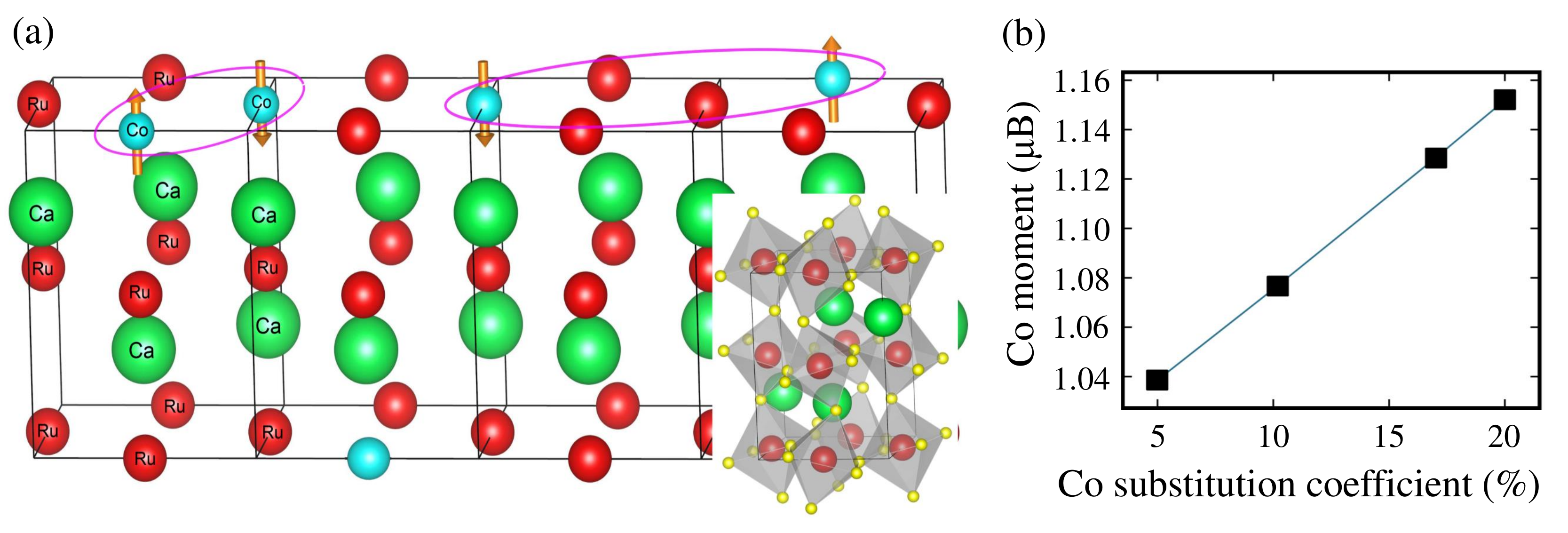} \vspace{-4mm}
\caption{(a) Schematic depiction of Ru substitution by Co-ion at random sites. Co ion assumes spin-1/2 configuration and interacts with neighboring and further away ions, forming entangled state. (b) Theoretically calculated moment of Co$^{4+}$ ion is found to be in the tune of $\sim$ 1 $\mu_B$.} \vspace{-6mm}
\end{figure*}

\textbf{Dimer formation and singlet-to-triplet transition in Ca(Co$_{x}$Ru$_{1-x}$)O$_{3}$}

The substitution of Ru ion by Co ion generates a random distribution of artificial spin-1/2 moments. As described schematically in Fig. 5, we have found that Co-substitution in x = 0.17 composition tends to create entangled exchange interaction pathways between neighboring and further away ions with ordered moment to the tune of $\sim$ 1 $\mu_B$. The artificial or quasi spin-1/2 of Co ion can develop liquid-like short-range dynamic correlation at low temperature. Elastic and inelastic neutron scattering measurements, performed on Multi-Axis Crystal Spectrometer (MACS) at the NIST Center for Neutron Research, revealed singlet-to-triplet transition between the entangled artificial spin-1/2 moments at low temperature of $T$ = 0.08 K, see Fig. 6. No magnetic order is detected in any of the stoichiometric compositions studied so far in Ca(Co$_{x}$Ru$_{1-x}$)O$_{3}$, 0$<$x$<$0.22. Fig. 6a shows the plot of background corrected and thermally balanced Q–E spectra at $T$ = 80 mK and in zero magnetic field. We observe the well-defined localized excitations at E = 5.9 meV at Q = 1 $\AA$$^{-1}$ and Q = 2 $\AA$$^{-1}$. The weak spectral weight at $Q$ = 2 $\AA$$^{-1}$ can be accounted for due to the Co form factor. It is important to point out that no such localized excitation was observed in CaRuO$_{3}$ or Ca(Co$_{x}$Ru$_{1-x}$)O$_{3}$ with low substitution coefficient, which further corroborates the dominant role of Co$^{4+}$- Co$^{4+}$ interaction in the gapped excitation. For quantitative analysis, we plot the averaged intensity of the inelastic spectra between $E$ = 4.5 and 7.5 meV as a function of wave-vector $Q$ at various temperatures in Fig. 6b. The width of the peaks is much broader than the instrument resolution, indicating the short-range nature of dynamic correlations. The integrated intensity vs. $Q$ is well described by the singlet-to-triplet transition,\cite{Mourigal} given by $I$(Q) = $\mid$F(Q)$\mid^{2}$ $\Sigma$ $m_{i}^{2}$[1 - sin(Q$d_{i}$)/(Q$d_{i}$)] where $d_i$, is the distance between neighboring ions and $m_i$ is ordered moment, at $T$ = 80 mK. The fitting involves not just the nearest neighbor spins, but also the next nearest neighbor spins. The estimated values of $d_i$, $d_1$ = 3.85 $\AA$ and $d_2$ = 5.4 $\AA$, are remarkably close to the actual lattice parameter values. It suggests that the collective excitation, indicating quantum fluctuation at low temperature, percolates beyond the nearest neighbors. Such behavior is typically observed in resonant valence bond candidate materials in spin liquid state.\cite{Lawler} However, in some cases, the disorder due to site switching impurity in RVB candidate materials has caused concern among the scientific community.\cite{Han} According to one school of thought, the quantum fluctuation at low temperature is attributed to the disorderness in two-dimensional frustrated lattice.

\textbf{Dispersive excitation at higher doping percentage in Ca(Co$_{x}$Ru$_{1-x}$)O$_{3}$}

To understand the evolution of spin dynamics in Ca(Co$_{x}$Ru$_{1-x}$)O$_{3}$ as a function of substitution coefficient x, we have further investigated the dynamic properties of Co$^{4+}$ ions in the insulating regime. Detailed inelastic neutron scattering measurements were performed on Ca(Co$_{0.2}$Ru$_{0.8}$)O$_{3}$ with x = 0.2. Unlike the observation of localized excitation in x = 0.17 compound, a dispersive magnetic excitation behavior is detected in x = 0.2 composition despite the absence of any magnetic order in the system. Plots of inelastic neutron data are shown in Fig. 7. As we can see in the figure, the dispersive excitation at low temperature of $T$ = 5 K tends to ebb as temperature increases. At $T$ = 150 K, the spin dynamics becomes much weaker with broad Q-dependence. It suggests that the correlated fluctuations, causing a dispersive excitation, breaks into weak short-range dynamic correlation as temperature increases.

\textbf{Quantum continuum fluctuation deep inside the spin glass state in optimally doped Ca(Co$_{0.15}$Ru$_{0.85}$)O$_{3}$}

The metal-insulator transition in cobalt-doped CaRuO$_{3}$ exhibits two distinct electrical transport regimes across the critical doping of x = 0.15. At x = 0.15, the system is found to manifest weak spin glass state below $T_G \sim$ 23 K with strong quantum continuum fluctuation at low temperature.\cite{Chen} Given the fact that the spin freezing and quantum fluctuation are at the two opposite extremes in a correlated electron system, the experimental observation is highly peculiar in nature. Magnetic measurements using ac susceptibility analysis technique exhibits weak ac frequency dependent cusp around $T_G$ =  23 K, see Fig. 8a-b. At $T$ $>$ 23 K, the static susceptibility depicts paramagnetic behavior, as a modest increase in $\chi$$^{'}$ is detected between $T$ = 300 K and 2 K. However, below $T \sim$ 23 K, the weak ac frequency dependent cusp in $\chi$$^{'}$, coupled with the concurrent frequency-dependent enhancement in $\chi$$^{"}$, hints of the glassiness in the system. Similar phenomena are often observed in spin glass type systems.\cite{Young} However, the spin glass state is not prominent, which could be attributed to the competing spin fluctuations in the system. The presence of spin glass state in Co-doped CaRuO$_{3}$ compound is also confirmed by multiple reports, albeit at different doping percentages.

\begin{figure}
\centering
\includegraphics[width=8.8 cm]{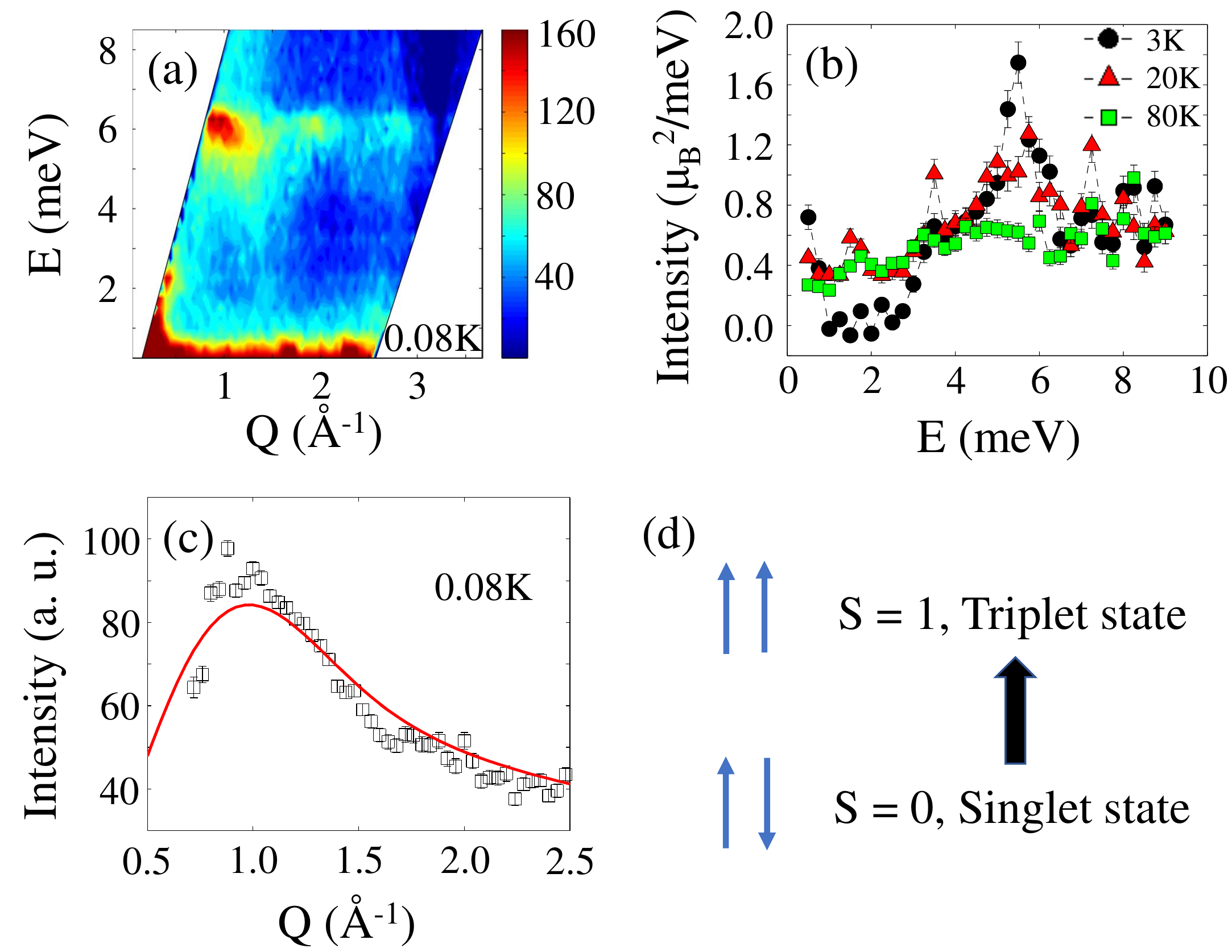} \vspace{-6mm}
\caption{(a) Energy–momentum (Q–E) map, depicting localized excitation in Ca(Co$_{0.17}$Ru$_{0.83}$)O$_{3}$, at $T$ = 0.08 K. (b) Energy scan at fixed $Q$ of 1.15 $\AA$$^{-1}$ at various temperatures. (c) Plot of integrated intensity between E = 4.5 meV and 7.5 meV vs Q. Experimental data is well described by the singlet-to-triplet transition due to the dimer-dimer interaction between nearest and further away spins, as described in Fig. 5a. (d) Schematic description of singlet to triplet transition.} \vspace{-6mm}
\end{figure}

The study of spin dynamics in Ca(Co$_{0.15}$Ru$_{0.85}$)O$_{3}$ is performed using inelastic neutron scattering measurements on high quality polycrystalline material. Fig. 8c-d show the $Q-E$ maps at $T$ = 1.5 K and $T$ = 300 K. Experimental data are background corrected and thermally balanced. At $T$ = 1.5 K, we notice that the entire energy-momentum space is occupied by the dynamic spectrum of the system. In fact, the system manifests a gapless continuous spin fluctuations to at least $E$ = 11.25 meV, which is the maximum accessible energy on MACS instrument at NIST. The gapless continuum excitation at low temperature is quantum mechanical in nature, as the broad spin fluctuation tends to subside as temperature increases, see Fig. 8e. At $T$ = 300 K, the dynamic behavior tends to disappear. It suggests that the spin dynamics of Co$^{4+}$ ions are local in nature that are reflected in the quantum continuum fluctuation. This argument is verified using the scaling analysis of dynamic susceptibility where the small scaling coefficient of $\alpha \sim$ 0.1 was inferred to reveal the extreme locality resulting in near Q-independence scaling.\cite{Chen} Most surprisingly, the continuum fluctuation coexists with the spin glass state in Ca(Co$_{0.15}$Ru$_{0.85}$)O$_{3}$. Unlike a conventional spin glass system where the spin freezing is well pronounced below the glass transition temperature, Ca(Co$_{0.15}$Ru$_{0.85}$)O$_{3}$ manifests weak glassy character. We argue that the strong spin fluctuations at low temperature significantly weakens the relaxation time of the nascent glassy phase and prohibits it from developing a static order. As we mentioned previously, no trace of any type of magnetic order is detected in the system. 

\begin{figure}
\centering
\includegraphics[width=8.8 cm]{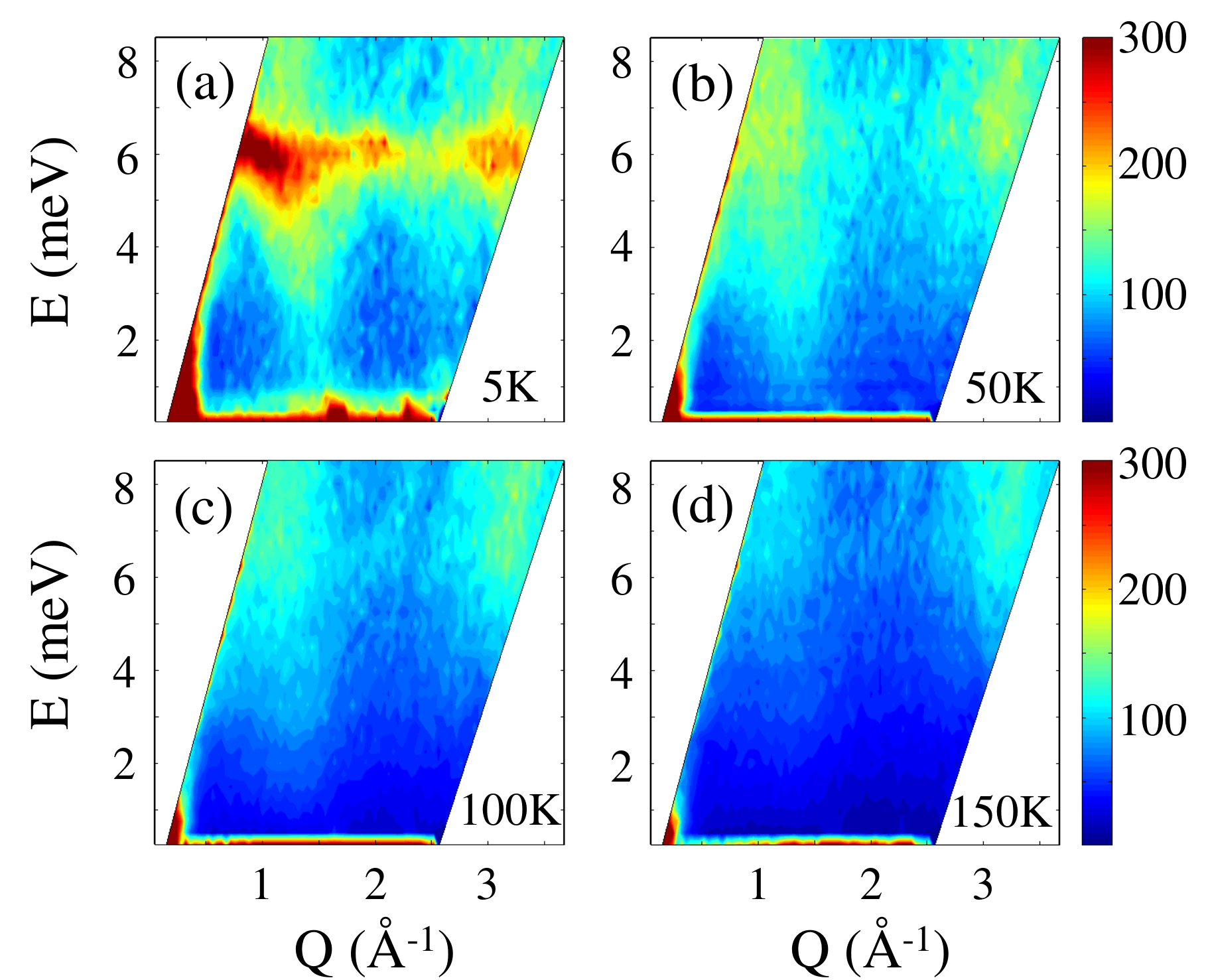} \vspace{-6mm}
\caption{(a) Q-E map obtained on MACS in Ca(Co$_{0.2}$Ru$_{0.8}$)O$_{3}$ compound at $T$ = 5 K. Localized excitation evolves into a dispersive behavior at higher substitution coefficient of x = 0.2. (b-c) The dispersive behavior weakens as temperature increases. (d) At $T$ = 150 K, weak remnants of spin dynamics with broad Q-dependences are detected.} \vspace{-6mm}
\end{figure}

\begin{figure*}
\centering
\includegraphics[width=15 cm]{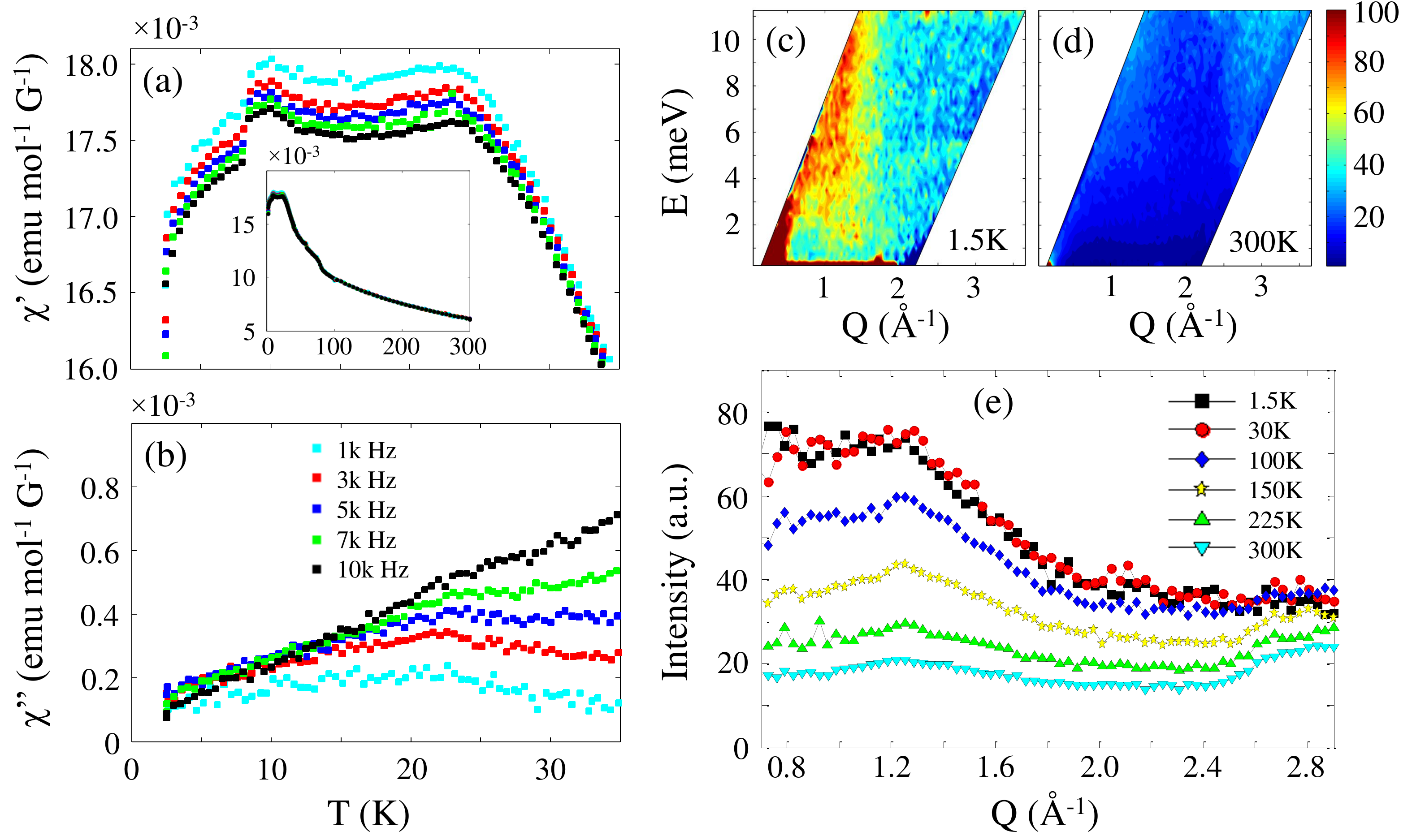} \vspace{-4mm}
\caption{(a-b) AC frequency dependent plots of static and dynamic susceptibilities in critically doped Ca(Co$_{0.15}$Ru$_{0.85}$)O$_{3}$ compound. A weak spin glass behavior is detected in this system. (c-d) Interestingly, a coexistence of glassy behavior with continuum fluctuation is detected in Ca(Co$_{0.15}$Ru$_{0.85}$)O$_{3}$ at low temperature. As temperature increases, the quantum continuum fluctuation subsides. (e) Plot of averaged intensity from $E$ = 2.5 to 9.5 meV vs. Q. Broad excitation in $Q$ is detected at low temperature. As temperature increases, neutron spectral weight becomes indistinguishable from the background.} \vspace{-4mm}
\end{figure*}

In summary, we have presented an overview of electrical and magnetic properties in stoichiometric and chemically doped CaRuO$_{3}$ perovskite. The ruthenate perovskite provides a rich platform to study many facets of quantum magnetism that are at the frontier of condensed matter physics research. In the stoichiometric composition, the parent compound exhibits a non-Fermi liquid behavior with complex magnetic ground state. The system is on the verge of ferromagnetic ordering. Yet, the dynamic nature of spin correlation prohibits the development of long-range magnetic order. Chemical substitution on Ru site by Ir and Co ions reveal metal-to-insulator transition as a function of the substitution coefficient. While the MIT process due to Ir substitution is found to be mediated by the underlying magnetic fluctuation, the Co substitution causes a host of new phenomena that resemble to the resonant valence bond physics. Most surprisingly, a coexistence of the weak spin glass phase with the quantum continuum state is detected in the critically doped Ca(Co$_{0.15}$Ru$_{0.85}$)O$_{3}$ compound. It can be argued that the disorder due to Co substitution be causing the glassy behavior and extreme locality, responsible for the continuum fluctuation, in the system. However, the same disorder, albeit at a nominally higher level, induces localized excitation, followed by the magnetic dispersive behavior at low temperature. Thus, the puzzling behavior of quantum continuum fluctuation in the spin glass state, which occurs in a material at the crossroad of Hubbard and Heisenberg formalism, poses new challenge to our understanding of quantum magnetism. A related question is, does the disorder play important role in quantum fluctuation. Perhaps, future research works on single crystal specimen can shed light on this pertinent problem.

The research at MU is supported by the Department of Energy, Office of Science, Office of Basic Energy Sciences under the grant no. DE-SC0014461. This work utilized neutron scattering facilities supported by Department of Commerce. Theoretical components of the research were performed by AE and VKD. The work of VKD was supported by the National Science Center in Poland.

\clearpage

\end{document}